\documentclass[fleqn,10pt]{wlscirep}
\usepackage[utf8]{inputenc}
\usepackage[T1]{fontenc}
\usepackage{amsmath}
\pdfoutput=1
\title{Understanding the Progression of Educational Topics via Semantic Matching}

\author[1,*]{Tamador Alkhidir}
\author[2,+]{Edmond Awad}
\author[3,+]{Aamena Alshamsi}
\affil[1]{Curriculum Department, Ministry of Education, United Arab Emirates}
\affil[2]{Department of Economics and Institute for Data Science and AI, University of Exeter, United Kingdom}
\affil[3]{Heuristic World, Dubai, United Arab Emirates}

\begin{abstract}

Education systems are dynamically changing to accommodate technological advances, industrial and societal needs, and to enhance students’ learning journeys. Curriculum specialists and educators constantly revise taught subjects across educational grades to identify gaps, introduce new learning topics, and enhance the learning outcomes. This process is usually done within the same subjects (e.g. math) or across related subjects (e.g. math and physics) considering the same and different educational levels, leading to massive multi-layer comparisons. Having nuanced data about subjects, topics, and learning outcomes structured within a dataset, empowers us to leverage data science to better understand the progression of various learning topics. In this paper, Bidirectional Encoder Representations from Transformers (BERT) topic modeling was used to extract topics from the curriculum, which were then used to identify relationships between subjects, track their progression, and identify conceptual gaps. We found that grouping learning outcomes by common topics helped specialists reduce redundancy and introduce new concepts in the curriculum. We built a dashboard to avail the methodology to curriculum specilists. Finally, we tested the validity of the approach with subject matter experts.

\end{abstract}
\begin{document}

\flushbottom
\maketitle
%
%
\thispagestyle{empty}

\section*{Introduction}

One of the most challenging aspects of curricular development is the identification of learning outcomes (LOs) which outline the concepts, knowledge and skills the students should acquire. Understanding and identifying relatedness of learning outcomes within and across subjects helps in preparing appropriate curricula, analyzing possible gaps, tutoring, advising students, designing adaptive learning systems, designing cross-curriculum activities, and supporting degree planning \cite{deng2010curriculum,nicholls2018developing,cunnington2014cultivating}.
Current curriculum alignment approaches rely heavily on manually labeling related learning outcomes. This practice is very tiresome, time-consuming, and requires continuous modification and updates to maintain various alignment objectives. This created a demand for curriculum mapping and alignment tools that can assess specialists to visualize curricular goals, scopes, sequences, and outcomes with added functionality such as filtering learning outcomes with certain words and tagging and relating learning outcomes \cite{holmes2018development,arafeh2016curriculum}. Yet, current curriculum-related activities still require high human intervention to map learning outcomes from different courses. Previous work tried to reduce the heavy dependence on subject domain experts (curriculum experts) by involving students in labeling the relatedness of different learning outcomes in courses. The initiative by MIT introduced a mechanism of crowdsourcing the identification of the relatedness of different subjects from students who attended the relevant courses \cite{miller2016crosslinks,huang2021network}. This approach can help in finding relationships between courses and highlighting some underlining misconceptions in students’ understanding; however, the resulted mapping is rigid and requires manual adjustment upon any change in course content. The advancement of natural language processing models and machine learning techniques can alleviate some of the previous limitations \cite{chary2019review}. For example, previous works reported the use of semantic matching techniques to relate various learning outcomes and create conceptual maps for a medical curriculum \cite{mondal2019learning}. However, suggested approaches used for supervised learning were semi-automated and required labeled training sets by subject domain experts in medicine. 

In this work, we analyse the relatedness between learning outcomes within and across subjects and educational levels and therefore draw connections between subjects in Ministry of education (MOE) curriculum Framework at United Arab Emirates (UAE). We use state-of-the-art semantic matching algorithms (BERT: Bidirectional Encoder Representation from Transformers)  \cite{devlin2018bert} to examine the relatedness between learning outcomes of 1-12 curriculum. Our methodology complements existing methods to write, verify and align learning outcomes. Our approach consists of the following process: The learning outcomes were converted into numerical representations using a BERT sentence transformer, and then dimension reduction techniques were used to identify the underlying common themes. By drawing the relationships between learning outcomes, we found that the related learning outcomes almost followed the general structure used by the MOE curriculum team, verifying the accuracy of our suggested methodology. The results were further analyzed, and different forms of visualization were generated to serve various purposes in curriculum planning including 1) checking the spirality of courses and how each learning outcome in a grade level is connected to the next grade and previous grade, 2) identifying the relatedness between different courses and 3) enhancing accessibility for educators, by presenting the findings in a user-friendly dashboard. Taken together, we offer a pipeline for applying a state-of-the-art language model algorithm in order to solve the problem of matching learning outcomes, for the first time (to our knowledge). 

\section*{Dataset and Case Study}
\subsection*{UAE curriculum Framework}

\begin{figure}[ht]
 \center
\includegraphics[width=0.6\textwidth]{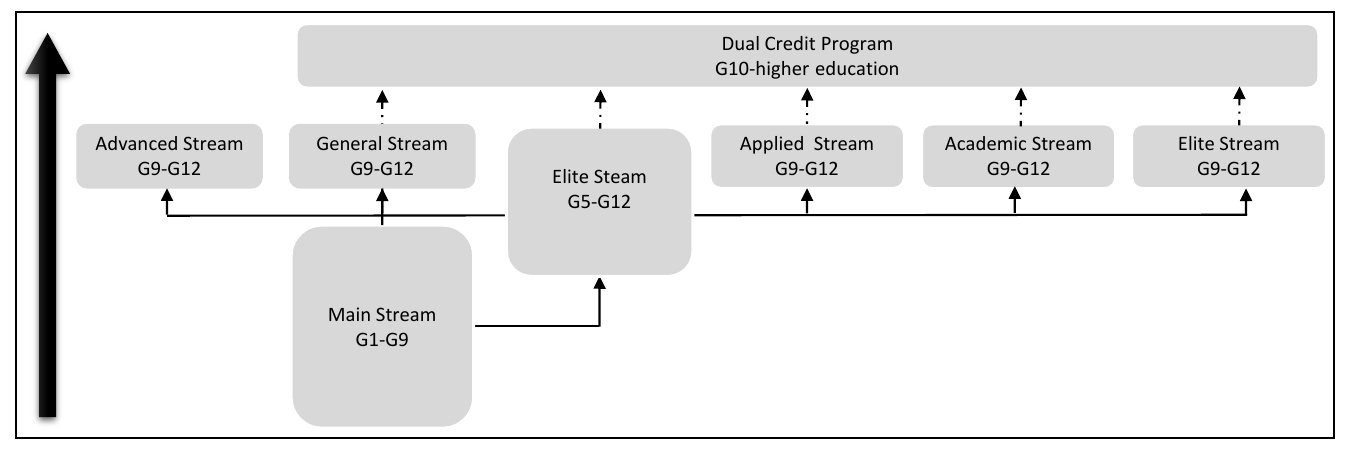}
\caption{ Description of the UAE educational system for 1-12 along with subjects covered by each stream}
\label{UAECu}
\end{figure}

Students in MOE curriculum are divided into multiple streams in alignment with their needs and interests (see Figure~\ref{UAECu}). The streaming starts from grade 5, where students are divided based on their performance, with the top 10\% of students in UAE admitted in the Elite stream and the rest in the main stream. The Elite stream continues until grade 12, while the main stream is divided into applied (vocational), advanced, general, and academics. The advanced and general streams address subjects with different difficulty levels mainly mathematics and science subjects. Generally, students are mandated to take group A subjects, whereas group B subjects can be optional depending on the stream.


In addition, students can take courses from university in replacement to some courses in tertiary education and vice versa in what is called the dual credit program. The dual credit program is still in the early phase of implementation. One main challenge in implementing the dual credit program is the fact that each university offers different learning outcomes for the same subject. Manually aligning all the university subjects with their corresponding subjects in tertiary education and ensuring correct alignment will be an exhausting job. Using advanced Natural Language Processing and AI tools can help in aligning university-level courses with tertiary courses to ensure there is no gap in students’ learning. Table~\ref{Table1} shows the courses included in this work and their abbreviation along with their designated stream. 


The MOE curriculum framework divides the subject into domains which in turn are divided into strands, standards, and learning outcomes in a hierarchical order.  Figure~\ref{strand} illustrates a snapshot from MOE curriculum framework for Computing, Creative Design and Innovation and an Integrated Science subject. The rest of the subjects follow the same structure with different statements for domains, strands, standards, and learning outcomes. Each layer addresses a more focused view of covered themes and topics in a subject. The learning outcomes layer contains most of the details needed to cover the pedagogical and topics aspects of the subject. Designing such a framework that addresses educational objectives and ensures a smooth educational journey for students requires massive time and effort. Curriculum planning is a rigorous process that involves setting themes and topics for each subject, depth of addressed concepts, and continuity of different topics across the grades. 

\subsection*{Learning Outcomes}
\begin{figure}[ht]
 \center
\includegraphics[width=0.6\textwidth]{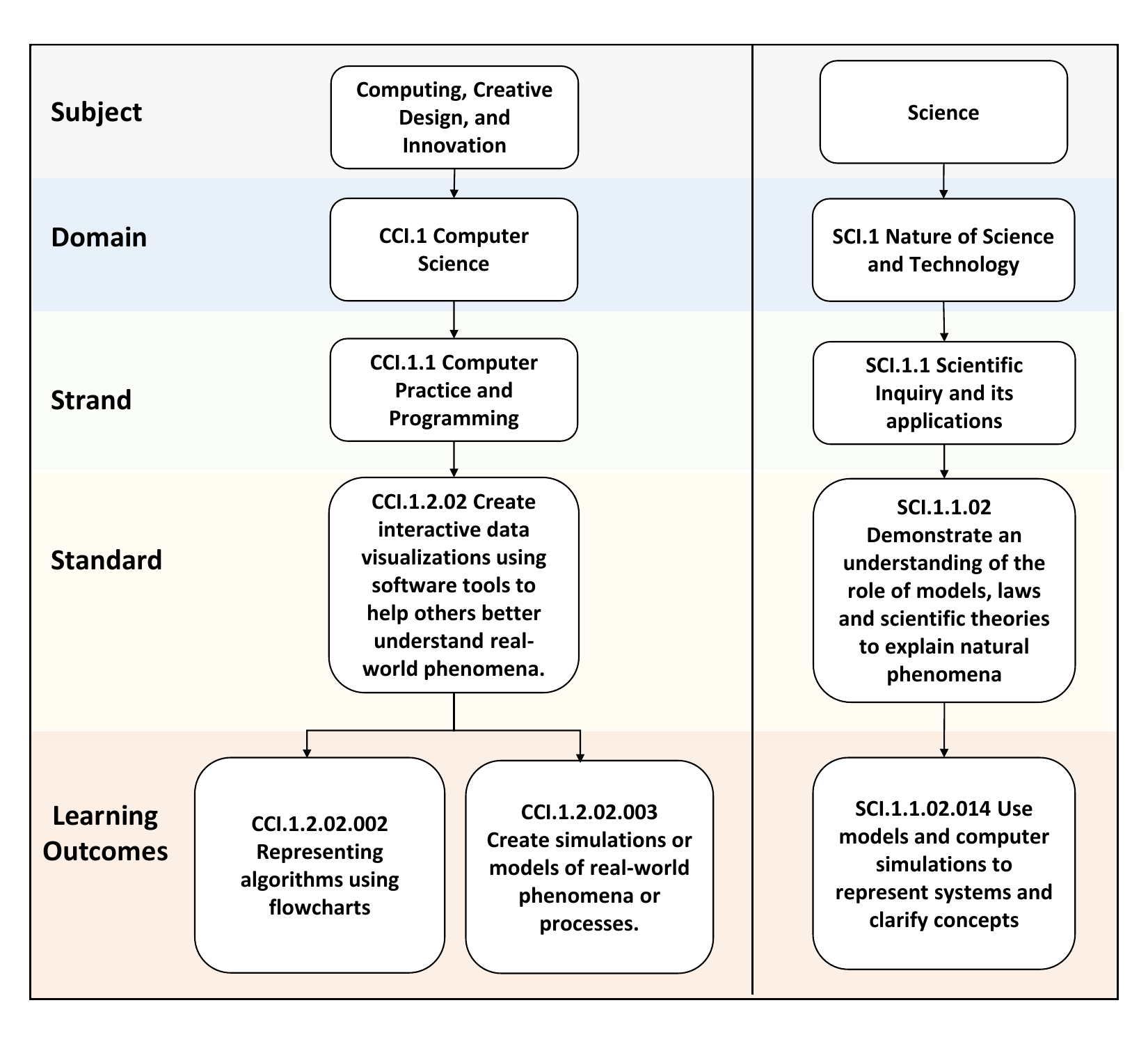}
\caption{ 1-12 UAE curriculum framework for computer science and science subject}
\label{strand}
\end{figure}

Automatic semantic matching between learning outcomes in various subjects can cut down the time needed to create an integrated curriculum. Common concepts and topics from different subjects can be automatically identified without the need to manually search for matches. This subsequently helps design better-integrated assessments and also checks the necessary prerequisites for each topic. Aligning subjects using the learning outcomes will provide more accurate results. For example, CCI.1.2.02.003  and CCI.1.2.02.002 have similar standards as shown in Figure ~\ref{strand}, but each learning outcome addresses partially different concepts with one being focused on algorithms and the other on building simulations, respectively. However, domain experts usually rely on the strands layer or standards layer to align multiple subjects as the number of learning outcomes is huge and it will be cumbersome to use the learning outcomes for mapping. In addition, the same concepts can be covered across multiple learning outcomes in different subjects. For example, CCI.1.2.02.002 in Computing, Creative Design and Innovation and SCI.1.1.02.014 in Integrated Science discussed the role of modeling and simulations in describing phenomena as indicated in Figure~\ref{strand}. Therefore, we have focused on the learning outcomes layer to align topics within and across grade levels to discover similar LOs that were not identified already. Since the similarity between LOs are not manually labeled, we used the framework to partially verify the efficacy of our methodology assuming LOs that belong to the same strands are similar. The number of learning outcomes included in this analysis is 7,43l. Thus, manually matching the learning outcomes is impractical as it requires checking around 55 million pairs of learning outcomes.

\section*{Methodology}

\subsection*{Language model transformers}
\begin{figure}[ht]
\center
\includegraphics[width=1\textwidth]{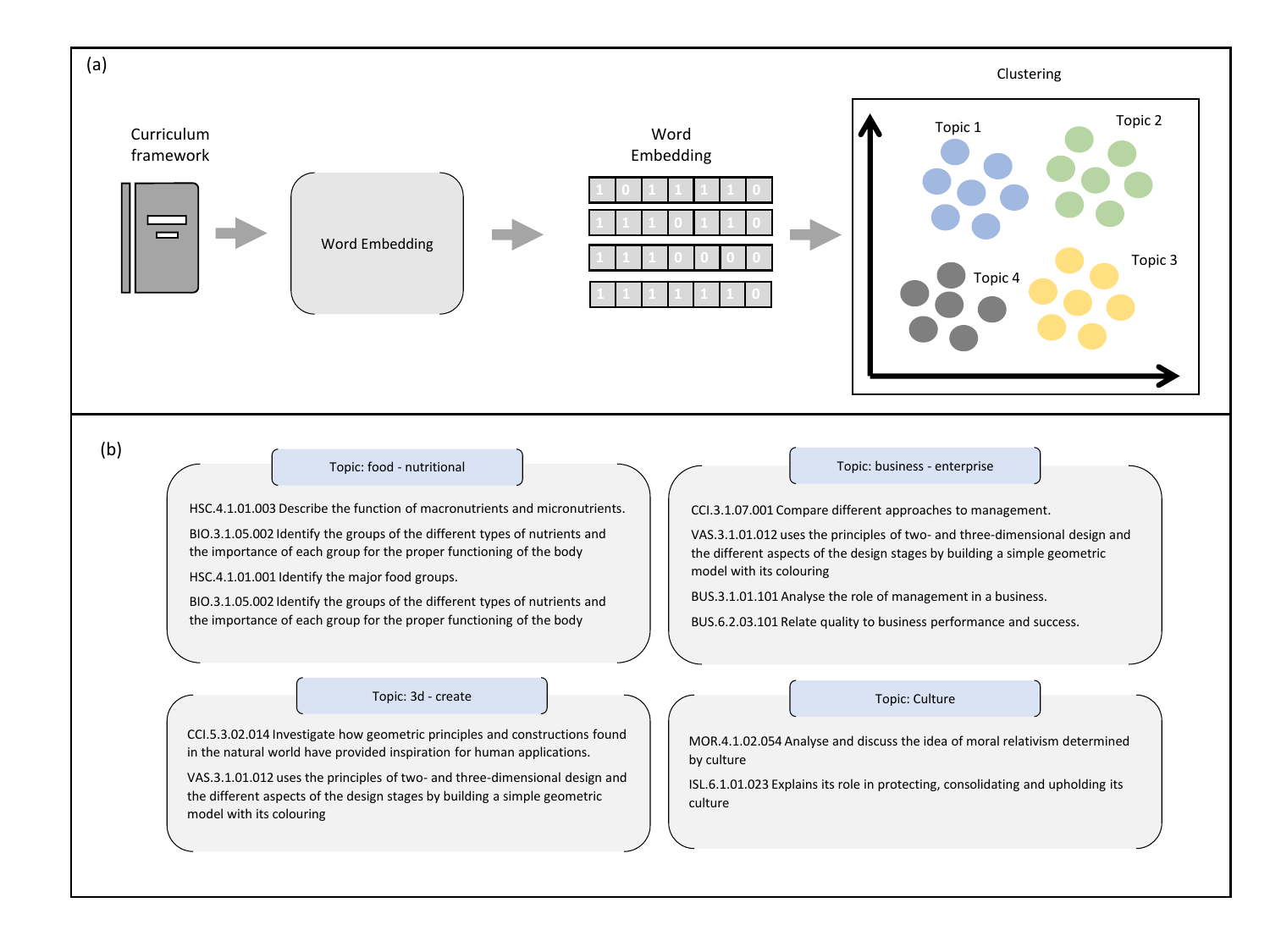}
\caption{Topic modeling is used to identify the main topics that are covered in different learning outcomes. Here are some examples of topics and their corresponding learning outcomes }
\label{Method}
\end{figure}

Language model transformers such as Bidirectional Encoder Representations from Transformers (BERT) have achieved remarkable results in neural machine translation, question answering, sentiment analysis, text summarization, and semantic similarity matching \cite{devlin2018bert}. Transformers were developed in replacement for traditional long short-term memory (LSTM) models that depend on passing and generating words sequentially, requiring more time to train, and producing less contextual language models\cite{sun2019patient}. Sentence transformers generate contextualized word embeddings considering the meaning of words in different sentences which differs from other conventional approaches that only consider the association between individual words (eg Word2Vec) \cite{rong2014word2vec}. Transformers can be coupled with other models to do more sophisticated language modeling, offering more scalability and adaptability to implement new applications. BERT is usually deployed by the pre-training deep neural network on labeled corpus and then fine-tunes the models depending on the task. The pre-training steps are done based on input tokenization, and the masked language model (MLM)\cite{mondal2019learning}. Masking specific words in the sentence and predicting the masked words by minimizing cross entropy are the main steps in MLM. BERT uses around 340 million parameters putting a high strain on computational resources \cite{devlin2018bert}. Knowledge distillation is an alternative lighter implementation of BERT; it is based on the original BERT model (teacher) training smaller models (students) and rewarding the winning models \cite{sanh2019distilbert}. This implementation cuts down the parameters by one order of magnitude, making the pre-trained model faster to run. Knowledge distillation was implemented using (“all-MiniLM-L12-v2”) sentence transformer \cite{wang2020minilm}. It maps sentences to a 384-dimensional dense vector space and is mostly used for clustering and semantic search. The “all-MiniLM-L12-v2” sentence transformer is a ready-to-use model that was already fine-tuned by finding the cosine similarity between 1 billion sentence pairs in a training corpus and applying cross-entropy loss with its correct pairs. Before feeding the learning outcomes to the sentence transformer, stop words, and numbers were removed from the learning outcomes. 
After that, the BERTtopic wrapper was used to assign a primary topic to each learning outcome. The model reduces the dimension of the vector space representation of the learning outcomes. This makes it easier to cluster the vectors and identify the underlying topics. The methodology is illustrated in Figure~\ref{Method}.a . The different learning outcomes were grouped into more than 400 topics such as nutrition, enterprise, 3D creation, and culture through Topic modeling as illustrated in Figure~\ref{Method}.b. A match between two learning outcomes is established when they share the same generated topic. 



\subsection*{Validation}
Since the similarity between LOs are not manually labeled, we used the current UAE curriculum framework to initially verify the efficacy of
our methodology assuming LOs that belong to the same strands are similar. 
To validate the effectiveness of the proposed method, we compared the matched pairs of learning outcomes detected by the proposed method with pairs of learning outcomes belonging to the same standard in the MOE curriculum framework depicted in Figure~\ref{UAECu} assuming LOs in the same standard are similar. We found that 7,072 learning outcomes (out of 7,431) with 95\% accuarcy belong to their corresponding standards as identified in the UAE curriculum framework. Although similarity between subjects is not manually labelled, we assume the existing framework can give us a good proxy of similarity to examine to efficacy of the suggested method. Failure to identify some existing similarity between LOs might be due to a limitation of the proposed method or a chance to uncover a similarity that was not addressed before.

To further validate the accuracy of our methodology for identifying related learning outcomes, we distributed 570 pairs of learning outcomes to specialists in 8 different subjects using two approaches. In the first approach, specialists were asked to identify matched pairs covering similar topics across two sets of subjects. The second approach involved assigning pairs of learning outcomes and asking specialists to categorize them as related or non-related. Both approaches were employed to ensure a balanced distribution of related and non-related learning outcomes. Out of the 570 LO pairs, 400 pairs (70\%)  were correctly labeled stating the possibility of using large language models in curriculum mapping. 


\section*{Results}
\subsection*{Subjects-Level Semantic Matching}
We use the number of LOs that have the same topic (generated by BERT) in different subjects to measure how similar those subjects are, across all educational stages. The matching percentage between the two subjects was calculated by normalizing the number of LOs in Subject A that matched LOs in Subject B ($\frac{ \textrm{Number of matched LOs in Subject A and Subject B}}{ \textrm{Total number of LOs in Subject A}}$).

\begin{figure}[ht]
\center
\includegraphics[width=1\textwidth]{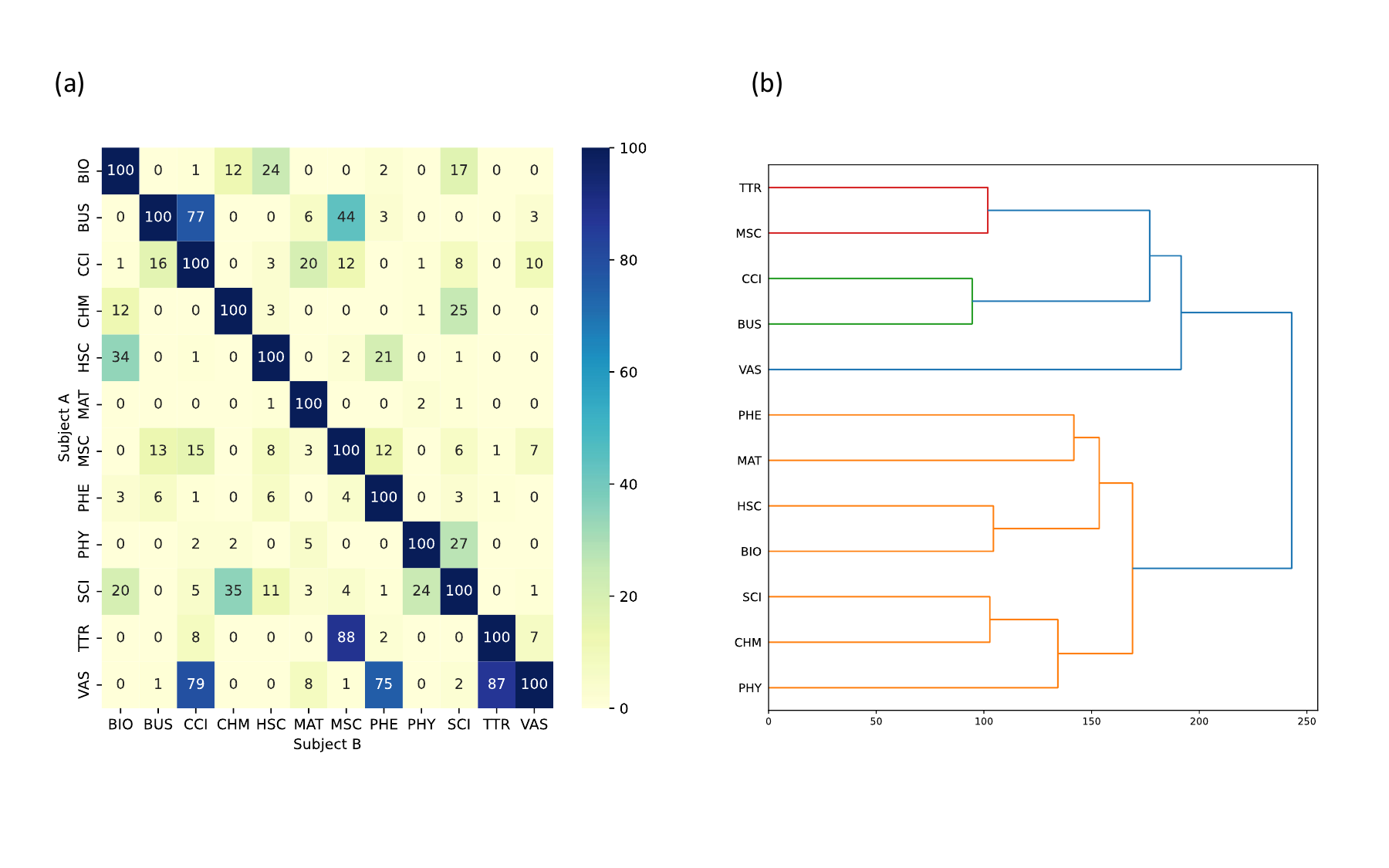}
\caption{ (a) The percentage of similarity between two subjects can be calculated by dividing the number of learning outcomes that they have in common by the total number of learning outcomes (b) Hierarchical clustering for the different subjects based on the calculated matching percentage }
\label{heatmap}
\end{figure}

Noteworthy, the similarity between the two subjects is asymmetric. Figure~\ref{heatmap} shows the matching percentage between different subjects spanning all grade levels.  Most of the subjects are interconnected with more than three subjects opening the opportunity for cross-curricula mapping and activities. The heatmap with hierarchical clustering depicted in Figure~\ref{heatmap} also serves the purpose of understanding relatedness between subjects. 
For example, Physics, Science, Biology, and Chemistry are closely clustered in Figure~\ref{heatmap}.b as they mostly address similar concepts in Science. In addition, the high overlap in learning outcomes (76\%) between Visual Arts (VAS) and Computing, Computing, Creative Design and Innovation (CCI) is primarily due to their shared focus on 3D design concepts. In addition, a significant portion (34\%) of Health Science and Biology learning outcomes overlapped, covering key areas like environmental concepts, nutrition and food, and physiology.

This demonstrates the capability of the methodology to identify new pairs of similar learning outcomes while reducing the time for curriculum specialists to compare a massive number of learning outcomes. The generated heatmap can guide specialists in reforming the curriculum framework to address cross-curricular concepts. For instance, additional concepts can be introduced to increase the connectivity between Mathematics, Business, and Computer Science as they often have interconnected concepts.

\subsubsection*{Conceptual Flow in Grade Level }
\begin{figure}[ht]
\center
\includegraphics[width=1\textwidth]{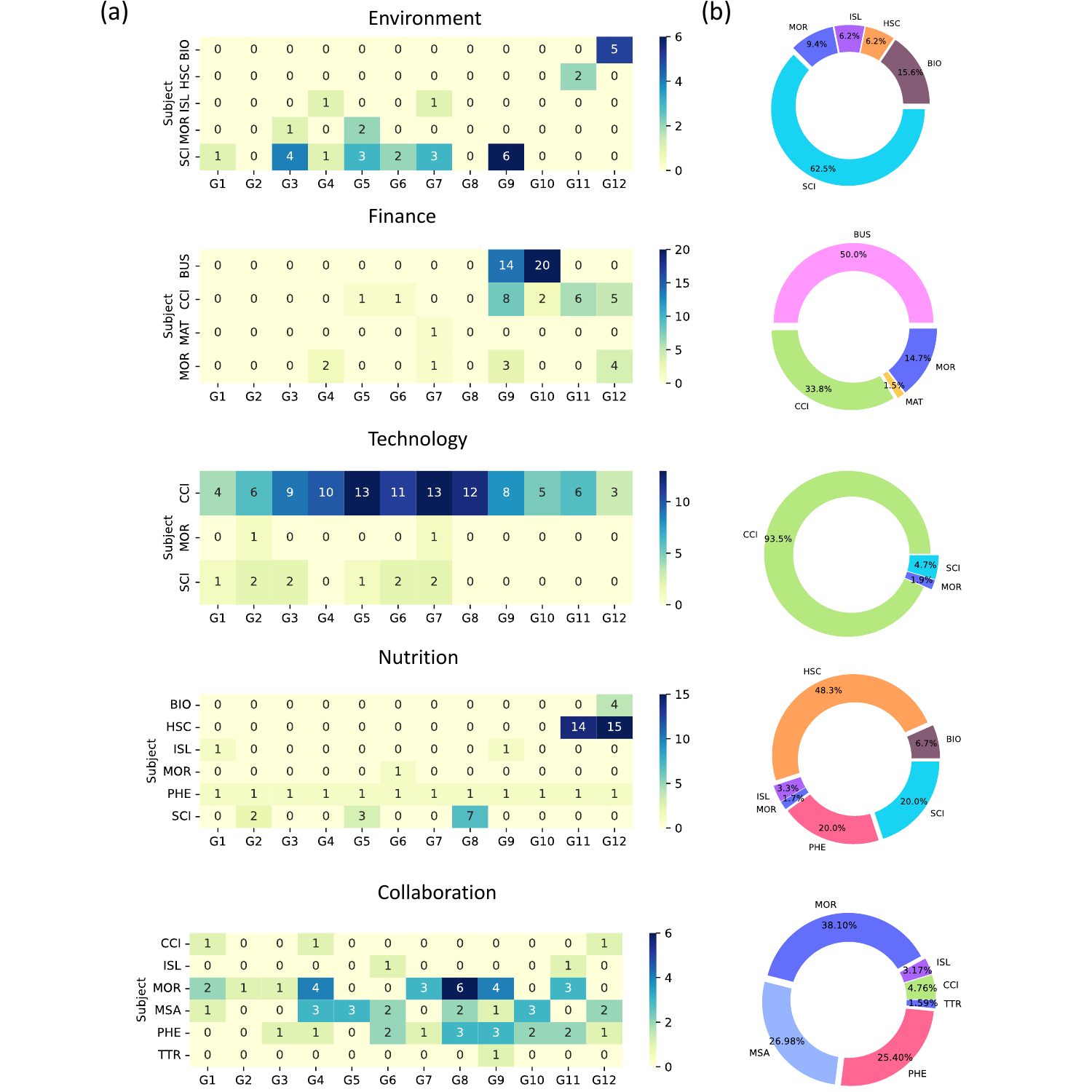}
\caption{ The most interconnected topics that are covered in more than three subjects and their distribution across grades 1-12. Environment topics are mostly discussed in science with few occurrences in other subjects. Finance as expected is discussed mostly in Business followed by Computing, Creative Design, and Innovation (CCI). Technology is mostly covered in CCI. The nutrition topic is covered in health science, then biology following physical science.}
\label{TopicDistribution}

\end{figure}

There are more than 20 topics generated from BERT topic modeling that appeared in various subjects as depicted in Figure \ref{TopicDistribution}. The learning outcomes labeled by these topics discussed multiple concepts in finance, nutrition, technology, and culture concepts. For example, concepts related to food and nutrition appear mostly in Health Science, then secondly in Physical Science then in Biology and Science, with a small appearance in Islamic study. Whereas, renewable energy appeared mostly in Science and secondly in  Computing, Creative Design and Innovation. 
The most important aspect of grouping learning outcomes with common topics is to understand whether learning outcomes follow the intended progression through the different grade levels (spirality). For example, learning outcomes associated with the topic "Environment" are addressed in most of the grade levels and mainly in Science, Moral Study, Islamic study, Health Science, and Biology. 
While the topic of "Nutrition" is extensively covered in Health Science programs, it receives minimal attention in Physical Science. This imbalance could be rectified by integrating nutrition topics within the Physical Science curriculum. 

This exercise will allow us to better plan the curriculum and determine whether a topic needs to be expanded in multiple grade levels. While understanding the appearance of concepts in different grade levels is important, it's equally crucial to analyze how they connect to a main concept and how skills are emphasized throughout the grades.

\begin{figure}[ht]
\center
\includegraphics[width=0.6\textwidth]{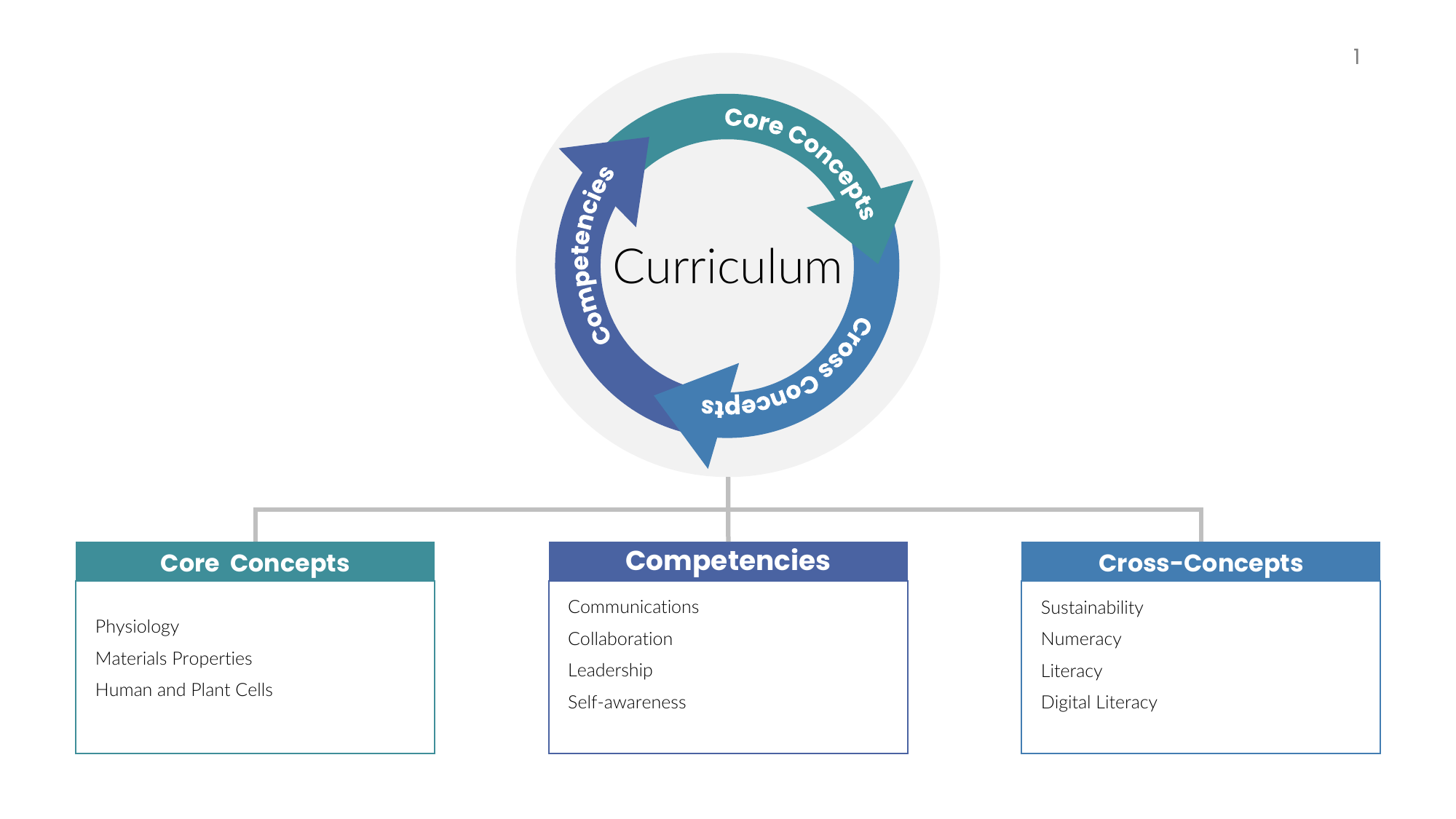}
\caption{Curriculum framework can be reshaped into three distinct components: core concepts, competencies, and cross-concepts, promoting a more holistic approach to learning}
\label{Cross}
\end{figure}

Therefore, current trends in curriculum design tend to divide the framework into core concepts, core competencies, and cross-concepts. Core concepts refer to the domains and themes that govern the subjects while other core competencies refer to the skills students need to acquire by the end of each grade level and cross-concepts are topics and competencies shared by multiple subjects as shown in Figure \ref{Cross}. The three-dimensional representation can help teachers better plan the lessons by understanding the key competencies and concepts that need to be covered. In addition, it helps the teacher to develop better strategies to enhance students' progression and accumulation of knowledge and skills.

Using the BERT topic, the current MOE framework shape can be transferred into the above components. For instance, collaboration as a core competency can be extracted from the different learning outcomes. The concepts of teamwork and collaboration appear mostly in the Moral Study as expected with fair distribution in multiple grades as illustrated in Figure\ref{TopicDistribution}. Such transformation would have a positive impact in reforming the curriculum to include practical skills that prepare students for job market.

\subsubsection*{Deployment}
\begin{figure}[ht]
\center
\includegraphics[width=1\textwidth]{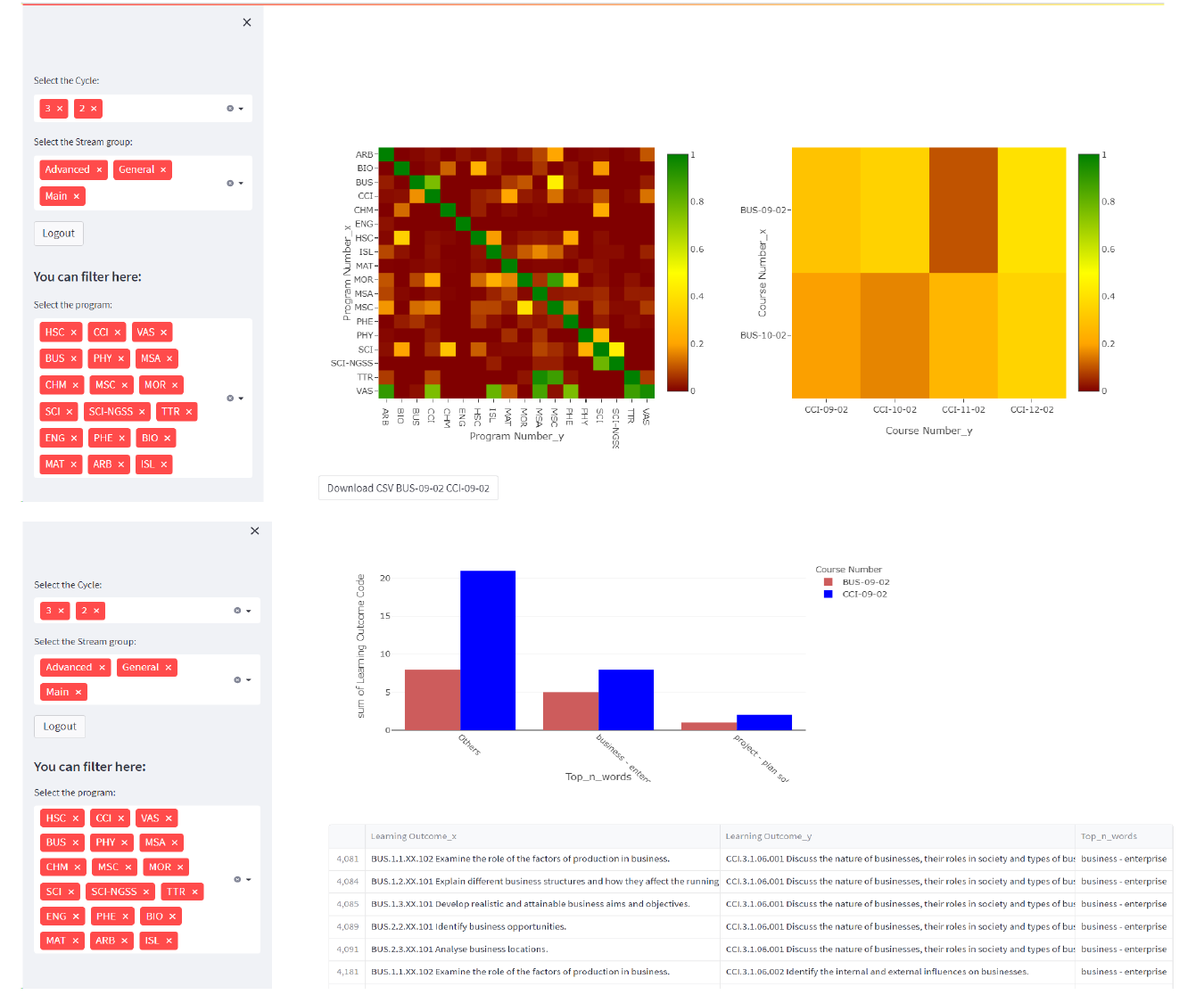}
\caption{Interactive dashboard for easy data exploration and filtering by stream and grade}
\label{Dashboard}
\end{figure}

An interactive dashboard was designed and built for the curriculum department using the open-source Streamlit package (https://github.com/streamlit/streamlit) to seamlessly search, navigate, and explore the learning outcomes and check similar ones. A heatmap was also a part of the interactive dashboard to guide specialists into the area of possible alignment. The dashboard allows specialists to filter the matched learning outcomes based on the courses, grade levels, streams, and percentage of matching (Figure~\ref{Dashboard}).  The dashboard is divided into the following parts:
\begin{itemize}
\item Filter section: Allow the user to filter according to the cycle, stream, and program (group of subjects at different grade levels
\item Program heat map: Matching percentage between different subjects visualized in a heat map
\item Subject heat map: Once a user clicks on a pair of subjects, another heat map will be generated to showcase the matching percentage across different grades.
\item Topic distribution: Bar chart indicating the number of learning outcomes per topic according to the selected pair of courses
\item Table: Present the learning outcome pairs under the same topic 

\end{itemize}

\section*{Conclusion}
In this paper, we propose an AI-based methodology to facilitate curriculum planning and save massive time and effort of manually aligning learning outcomes within and across subjects. We show that semantic matching can group similar learning outcomes within and across subjects and achieve results that generally match current curriculum structure. We design a data science pipeline to further analyze detected similarities to cluster topics and identify dependencies. Our methodology demonstrates interesting similarities that were not captured previously between courses. We made the findings available to MOE curriculum experts in a dashboard to further zoom in and examine cross-mapping between courses. The resultant semantic matching allows specialists to check for spirality within a subject across multiple grade levels, and check common interconnected topics that govern the curriculum.
For future implementation, we can deduce learning outcomes and keywords from lessons in textbooks using a state-of-the-art predefined summarization sentence transformer and link lessons across different subjects automatically. In this paper, we focus only on the semantic aspect of the learning outcomes, but we can additionally link learning outcomes through cognitive skills (bloom taxonomy levels) to address the complexity of the learning outcomes~\cite{krathwohl2002revision}. Furthermore, we can deduce prerequisites by combining semantic matching results with students’ performance in outcomes-based assessments and pinpoint how not understanding one learning outcome can affect others.

%


\bibliography{Ref}

\section*{Acknowledgements}
The author would like to thank the curriculum department at
Ministry of Education led by Dr. Rabaa Alsumaiti in the United Arab Emirates for providing 
the data and articulating the problem statement A special thanks to
Hamdan AlShakeili for his massive help in configuring spark






\begin{table}[ht]
\center
\caption{The abbreviation, name, and type of all subjects in the 1-12 UAE system. The abbreviations are used later in the paper in different plots and results.}\label{Table1}
\resizebox{0.5\textwidth}{!}{%
\begin{tabular}{@{}lll@{}}
\toprule
Abbreviation & Subject Name                           & Subject Type  \\ \midrule
BIO          & Biology                                & Group A       \\
BUS          & Business Studies                       & Group B       \\

CHM          & Chemistry                              & Group A       \\
CCI          & Computing, Creative Design and Innovation                    & Group B       \\
HSC          & Health Science                         & Group B       \\
ISL          & Islamic study                        & Group A       \\
MAT          & Mathematics                            & Group A       \\
MSA          & Music                            & Group A       \\
MOR          & Moral Study                            & Group A       \\
PHE          & Physical Education                     & Group B       \\
PHY          & Physics                                & Group A       \\
SCI          & Integrated Science                     & Group A       \\
SST          & Social Study                     & Group A       \\
TTL          & Applied Travel, Tourism, and Leisure    & Applied       \\ 
VAS          & Visual Art   & Group A       \\ \bottomrule
\end{tabular}%
} 
\end{table}

\end{document}